\documentclass[12pt]{iopart}
\usepackage{amsfonts}
\usepackage{amsbsy}

\begin{document}

\renewcommand\vec[1]{\boldsymbol{#1}}
\newcommand{\myfrac}[2]{\frac{\displaystyle #1}{\displaystyle #2}}

\title{Lattice representation and dark solitons of the Fokas-Lenells equation.}
\author{V.E. Vekslerchik$^{1,2}$}

\address{$^{1}$
Institute for Radiophysics and Electronics, Kharkov, Ukraine
}

\address{$^{2}$
Universidad de Castilla-La Mancha, Ciudad Real, Spain
}

\ead{vekslerchik@yahoo.com}

\begin{abstract}
This work is devoted to an integrable generalization of the nonlinear Schr\"odinger
equation proposed by Fokas and Lenells. I discuss the relationships between this
equation and other integrable models. Using the reduction of the Fokas-Lenells equation to the
already known ones I obtain the N-dark soliton solutions.
\end{abstract}

\pacs{02.30.Ik, 02.30.Jr, 45.05.+x}

\submitto{\NL} 

\maketitle

\section{Introduction.}

This paper is devoted to the Fokas-Lenells equation (FLE) 
\cite{F1995,L2009,LF2009}, 
\begin{equation}
  \left\{ 
  \begin{array}{lcl}
  u_{xy} + u - 2i uv u_{x} & = & 0
  \\[1mm]
  v_{xy} + v + 2i uv v_{x} & = & 0
  \end{array}
  \right.
\label{lfe}
\end{equation}
which appears in physical applications in the form
\begin{equation}
  u_{xy} + u \mp 2i |u|^{2} u_{x} = 0.
\end{equation}
FLE is known (see \cite{F1995}) to be an integrable equation
describing the first negative flow of the integrable hierarchy associated with 
the derivative nonlinear Schr\"odinger (DNLS) equation.
In papers \cite{L2009,LF2009} the authors derived the FLE from the Maxwell equations 
for the propagation of femtosecond pulses in optical fibers, 
established its bi-Hamiltonian structure, 
elaborated the inverse scattering transform scheme for the case of zero 
boundary conditions and 
obtained the N-bright soliton solutions.
The aim of this paper is two-fold.
Firstly, I want to derive the N-dark soliton solutions. 
Secondly, I am going to establish the relationships between the FLE and 
other integrable models: the nonlinear Schr\"odinger (NLS) equation, 
the Merola-Ragnisco-Tu (MRT) equations \cite{MRT1984} 
and the Ablowitz-Ladik (AL) model \cite{AL1975}.

It this paper I present the commuting pair of the B\"acklund 
transformations (section \ref{sec-bt}) which I use in section 
\ref{sec-lattice} to obtain the lattice representation of the FLE. 
After bilinearizing the lattice equations and establishing their 
relationship with the AL hierarchy  
I use the N-soliton solutions of the latter to obtain 
the N-soliton solutions for the FLE (section \ref{sec-solitons}). 
Finally, in section \ref{sec-intersections} I discuss the links between 
the DNLS, NLS, MRT and AL equations.

\section{B\"acklund transformations. \label{sec-bt}}

The method of this work is based on two transformations, 
$\hat\mathcal{T}$ and $\check\mathcal{T}$, given by 

\begin{equation}
  \hat\mathcal{T}: \qquad (u,v) \to \left( \hat{u}, \hat{v} \right), \qquad
  \left\{
  \begin{array}{lcl}
  \hat{v} & = & 
  - i v_{y} + uv^{2}
  \\
  \hat{u} & = & 
  \myfrac{i}{\hat{v}_{x}} 
  - \myfrac{1}{v}
  \end{array}
  \right.
\label{transform-1}
\end{equation}
and
\begin{equation}
  \check\mathcal{T}: \qquad (u,v) \to \left( \check{u}, \check{v} \right), \qquad
  \left\{
  \begin{array}{lcl}
  \check{u} & = & i u_{y} + u^{2} v 
  \\
  \check{v} & = & 
  \myfrac{ 1 }{ i \check{u}_{x}} 
  - \myfrac{ 1 }{ u }. 
  \end{array}
  \right.
\label{transform-2}
\end{equation}
These transformations possess two important properties:

\begin{itemize}
\item
Transformations $\hat\mathcal{T}$ and $\check\mathcal{T}$ 
send solutions of the FLE to solutions of the FLE.

\item
Transformations $\hat\mathcal{T}$ and $\check\mathcal{T}$ are inverse 
when applied to solutions of the FLE.

\end{itemize}

To prove the first statement consider 
the quantities $\Delta[u,v]$ and $\bar\Delta[u,v]$ 
which are the left-hand sides of equations (\ref{lfe}):
\begin{equation}
  \begin{array}{lcl}
  \Delta[u,v] & = & u_{xy} + u - 2i uv u_{x} 
  \\[1mm]
  \bar\Delta[u,v] & = & v_{xy} + v + 2i uv v_{x} 
  \end{array}
\end{equation}
It can be shown by direct calculations that if pairs $(u,v)$ and 
$(\hat{u},\hat{v})$ are related by (\ref{transform-1}), then
\begin{eqnarray}
  \Delta[ \hat{u}, \hat{v} ] & = & 
  \frac{ \bar\Delta[u,v] }{ v^{2} } 
  - i \frac{\partial}{\partial x} 
      \frac{ \bar\Delta\left[ \hat{u},\hat{v} \right] }{ \hat{v}_{x}^{2} } 
\\
  \bar\Delta[\hat{u}, \hat{v} ] & = & 
  v^{2} \Delta[u,v] - 
  \left( i \frac{\partial}{\partial y} + \frac{ 2 \hat{v} }{ v } \right)
  \bar\Delta[u,v]. 
\end{eqnarray}
From these equations one can easily see that if $u$ and $v$ are solutions 
of the FLE, $\Delta[u,v] = \bar\Delta[u,v] = 0$, then 
$\Delta[\hat{u},\hat{v}] = \bar\Delta[\hat{u},\hat{v}] = 0 $ 
which means that $\hat{u}$ and $\hat{v}$ also satisfy the FLE.
In a similar way one can show 
that $\Delta$ and $\bar\Delta$ calculated for the 
$\check\mathcal{T}$-transformed functions are given by
\begin{eqnarray}
  \Delta[ \check{u}, \check{v} ] & = & 
  \left( i \frac{\partial}{\partial y} - \frac{ 2 \check{u} }{ u } \right)
    \Delta[u,v]
  + u^{2} \bar\Delta[u,v] 
\\
  \bar\Delta[ \check{u}, \check{v} ] & = & 
  \frac{ \Delta[ u, v ] }{ u^{2} } 
  + i \frac{\partial}{\partial x} 
      \frac{ \Delta\left[ \check{u}, \check{v} \right] }{ \check{u}_{x}^{2} }. 
\end{eqnarray}
Thus the identities $\Delta[u,v] = \bar\Delta[u,v] = 0$ lead to 
$\Delta[\check{u},\check{v}] = \bar\Delta[\check{u},\check{v}] = 0 $.
To summarize,
\begin{equation}
  \Delta[u,v] = \bar\Delta[u,v] = 0
  \qquad\Rightarrow\qquad
  \left\{
  \begin{array}{l}
  \Delta[\hat{u},\hat{v}] = \bar\Delta[\hat{u},\hat{v}] = 0
  \\[1mm]
  \Delta[\check{u},\check{v}] = \bar\Delta[\check{u},\check{v}] = 0
  \end{array}
  \right.
\label{all-deltas}
\end{equation}
which proves the fact that transformations 
$\hat\mathcal{T}$ and $\check\mathcal{T}$ 
send solutions of the FLE to solutions of the FLE.
 
To demonstrate that 
\begin{equation}
  \hat\mathcal{T} \circ \check\mathcal{T} = 
  \check\mathcal{T} \circ \hat\mathcal{T} = 
  1 \; \mbox{mod} \left( \Delta, \bar\Delta \right)
\label{lemma-2-f}
\end{equation}
consider the results of joint action of 
$\hat\mathcal{T}$- and $\check\mathcal{T}$-transformations:
\begin{equation}
  \pmatrix{ \tilde{u} \cr \tilde{v} } = 
  \hat\mathcal{T} \check\mathcal{T} \pmatrix{ u \cr v }, 
  \qquad
  \left\{
  \begin{array}{lcl}
  \tilde{v} & = & 
  - i \check{v}_{y} + \check{u}\check{v}^{2}
  \\
  \tilde{u} & = & 
  \myfrac{i}{\tilde{v}_{x}} 
  - \myfrac{1}{ \check{v} }
  \end{array}
  \right.
\label{transform-12}
\end{equation}
and
\begin{equation}
  \pmatrix{ \breve{u} \cr \breve{v} } = 
  \check\mathcal{T} \hat\mathcal{T} \pmatrix{ u \cr v }, 
  \qquad
  \left\{
  \begin{array}{lcl}
  \breve{u} & = & i \hat{u}_{y} + \hat{u}^{2} \hat{v} 
  \\
  \breve{v} & = & 
  \myfrac{ 1 }{ i \breve{u}_{x}} 
  - \myfrac{ 1 }{ \hat{u} }, 
  \end{array}
  \right.
\label{transform-21}
\end{equation}
where $\hat{u}$, $\hat{v}$, $\check{u}$ and $\check{v}$ are defined in 
(\ref{transform-1}), (\ref{transform-2}).
Using definitions (\ref{transform-2}) one can obtain from (\ref{transform-12}) 
\begin{equation}
  \tilde{u} - u = 
  \frac{ u }{ \check{u}_{x} \check{v} \tilde{v}_{x} } \;
  \bar\Delta[ \check{u}, \check{v} ],
  \qquad
  \tilde{v} - v = 
  \frac{ 1 }{ \check{u}_{x}^{2} } \;
  \Delta[ \check{u}, \check{v} ].
\end{equation}
In a similar way definitions (\ref{transform-21}) and (\ref{transform-1}) 
lead to 
\begin{equation}
  \breve{u} - u = 
  \frac{ 1 }{ \hat{v}_{x}^{2} } \;
  \bar\Delta[ \hat{u}, \hat{v} ],
  \qquad
  \breve{v} - v = 
  \frac{ v }{ \hat{u} \hat{v}_{x} \breve{u}_{x} } \;
  \Delta[ \hat{u}, \hat{v} ].
\end{equation}
These identities together with (\ref{all-deltas}) prove (\ref{lemma-2-f})

In what follows I will use the symbol $\mathcal{T}$ for both transformations, 
\begin{equation}
  \hat\mathcal{T} = \mathcal{T},
  \qquad
  \check\mathcal{T} = \mathcal{T}^{-1}
\end{equation}
bearing in mind that $\mathcal{T}$-transformations are used only with  
solutions of the FLE.

\section{Lattice representation of the FLE and bilinearization. 
\label{sec-lattice}}

Using the transformations discussed in the previous section I introduce 
(instead of single solution of the FLE, $u$ and $v$) an infinite sequence 
of solutions $u_{n}$ and $v_{n}$ defined by iteration of 
$\mathcal{T}$-transformations:
\begin{equation}
  \pmatrix{ u_{n} \cr v_{n} } = 
  \mathcal{T}^{n} \pmatrix{ u \cr v },
  \qquad
  n= 0, \pm 1, \pm 2, ...
\end{equation}
In terms of  $u_{n}$ and $v_{n}$ the definitions of transformations 
$\mathcal{T}$ and $\mathcal{T}^{-1}$, 
equations (\ref{transform-1}) and (\ref{transform-2}), can be written as 
\begin{equation}
  \left\{
  \begin{array}{lcl}
  - i \partial_{x} v_{n} & = & 
    \myfrac{ v_{n-1} }{ 1 + u_{n}v_{n-1} }
  \\[2mm] 
  - i \partial_{y} v_{n} & = & 
    v_{n+1} - u_{n}v_{n}^{2}
  \end{array}
  \right.
\end{equation}
and 
\begin{equation}
  \left\{
  \begin{array}{lcl}
  i \partial_{x} u_{n} & = & 
    \myfrac{ u_{n+1} }{ 1 + u_{n+1}v_{n} }
  \\[2mm] 
  i \partial_{y} u_{n} & = & 
    u_{n-1} - u_{n}^{2}v_{n}.
  \end{array}
  \right.
\end{equation}
It is easy to see that the above equations can be regrouped in two systems
\begin{equation}
  \left\{
  \begin{array}{rcl}
  i \partial_{y} u_{n} & = & 
    u_{n-1} - u_{n}^{2}v_{n}
  \\
  - i \partial_{y} v_{n} & = & 
    v_{n+1} - u_{n}v_{n}^{2}
  \end{array}
  \right.
\label{MRT-eq-1}
\end{equation}
and
\begin{equation}
  \left\{
  \begin{array}{rcl}
  i \partial_{x} u_{n} & = & 
    \myfrac{ u_{n+1} }{ 1 + u_{n+1}v_{n} }
  \\[2mm] 
  - i \partial_{x} v_{n} & = & 
    \myfrac{ v_{n-1} }{ 1 + u_{n}v_{n-1} }
  \end{array}
  \right.
\label{MRT-eq-2}
\end{equation}
These equations are not new: they belong to the MRT hierarchy 
that has been introduced, in the framework of the reduction technique for 
Poisson-Nijenhuis structures, in \cite{MRT1984} as an integrable lattice 
hierarchy, whose continuum limit is the AKNS hierarchy.
In \cite{MRT1984} the authors demonstrated that this hierarchy is endowed with 
a canonical Poisson structure and admits a vector generalisation.
They solved the associated spectral problem and explicity contructed 
action-angle variables through the $r$-matrix approach.
In this work I show  that equations (\ref{MRT-eq-1}) and (\ref{MRT-eq-2}), 
as well as the whole MRT hierarchy are closely related to 
another discrete version of the AKNS equations, namely the famous AL hierarchy.
To this end I bilinearize equations (\ref{MRT-eq-1}) and (\ref{MRT-eq-2}) by 
introducing the $\tau$-functions $\rho_{n}$, $\sigma_{n}$ and $\tau_{n}$ 
related by 
\begin{equation}
  \tau_{n}^{2} - \tau_{n-1}\tau_{n+1} = \rho_{n}\sigma_{n} 
\label{tau-restiction}
\end{equation}
as follows:
\begin{equation}
  u_{n} = \frac{ \sigma_{n-1} }{ \tau_{n} }, 
  \qquad
  v_{n} = \frac{ \rho_{n} }{ \tau_{n-1} }. 
\label{uv-bilin}
\end{equation}
By straightforward algebra one ca obtain that
\begin{eqnarray}
  i \partial_{x} u_{n} - \frac{ u_{n+1} }{ 1 + u_{n+1}v_{n} } & = & 
  \frac{ 1 }{ \tau_{n-1} \tau_{n} }
  \left[ 
    \mathcal{I}^{(+)}_{n-1} 
    - u_{n} \mathcal{J}_{n-1} 
  \right] 
\\[2mm]
  -i \partial_{x} v_{n} - \frac{ v_{n-1} }{ 1 + u_{n}v_{n-1} } & = & 
  \frac{ 1 }{ \tau_{n-1} \tau_{n} }
  \left[ 
      \mathcal{I}^{(-)}_{n} 
    - v_{n} \mathcal{J}_{n-1} 
  \right] 
\\[2mm]
  i \partial_{y} u_{n} - u_{n-1} + u_{n}^{2}v_{n}  & = & 
  \frac{ 1 }{ \tau_{n-1} \tau_{n} } 
  \left[ \mathcal{K}^{(+)}_{n-1} - u_{n} \mathcal{L}_{n-1} \right]
\\[2mm]
  - i \partial_{y} v_{n} - v_{n+1} + u_{n}v_{n}^{2}  & = & 
  \frac{ 1 }{ \tau_{n-1} \tau_{n} } 
  \left[ \mathcal{K}^{(-)}_{n} - v_{n} \mathcal{L}_{n-1} \right]
\end{eqnarray}
where
\begin{eqnarray}
  \mathcal{I}^{(+)}_{n} & = & 
  i D_{x} \, \sigma_{n} \cdot \tau_{n} - \tau_{n-1} \sigma_{n+1} 
\\ 
  \mathcal{I}^{(-)}_{n} & = & 
  i D_{x} \, \tau_{n} \cdot \rho_{n} - \rho_{n-1} \tau_{n+1} 
\\ 
  \mathcal{J}_{n} & = & 
  i D_{x} \, \tau_{n+1} \cdot \tau_{n} + \rho_{n} \sigma_{n+1} 
\\ 
  \mathcal{K}^{(+)}_{n} & = & 
  i D_{y} \, \sigma_{n} \cdot \tau_{n} - \sigma_{n-1} \tau_{n+1} 
\\ 
  \mathcal{K}^{(-)}_{n} & = & 
  i D_{y} \, \tau_{n} \cdot \rho_{n} - \tau_{n-1} \rho_{n+1} 
\\ 
  \mathcal{L}_{n} & = & 
  i D_{y} \, \tau_{n+1} \cdot \tau_{n} - \sigma_{n} \rho_{n+1} 
\end{eqnarray}
and $D_{x}$ and $D_{y}$ are the Hirota's bilinear operators defined as
\begin{equation}
  D_{x}^{k} \, D_{y}^{l} \, a(x,y) \cdot b(x,y) = 
  \left.
  \frac{ \partial^{k+l} }{ \partial \xi^{k} \partial \eta^{l} } \; 
  a(x + \xi, y + \eta) \, b(x - \xi, y - \eta) 
  \right|_{\xi=\eta=0}.
\end{equation}
One can easily see that equations (\ref{MRT-eq-1}) and (\ref{MRT-eq-2}) can be 
satisfied by imposing the conditions
\begin{equation}
  \mathcal{I}^{(\pm)}_{n} = 
  \mathcal{J}_{n} =  
  \mathcal{K}^{(\pm)}_{n} = 
  \mathcal{L}_{n}  = 0. 
\end{equation}
Thus, instead of system (\ref{MRT-eq-1}) and (\ref{MRT-eq-2}) one can 
solve the bilinear system given by
\begin{eqnarray}
  i D_{x} \, \sigma_{n} \cdot \tau_{n} - \tau_{n-1} \sigma_{n+1} 
  & = & 0 
\label{eq-bilin-1}
\\ 
  i D_{x} \, \tau_{n} \cdot \rho_{n} - \rho_{n-1} \tau_{n+1} 
  & = & 0 
\\ 
  i D_{x} \, \tau_{n+1} \cdot \tau_{n} + \rho_{n} \sigma_{n+1} 
  & = & 0 
\label{eq-bilin-3}
\\ 
  i D_{y} \, \sigma_{n} \cdot \tau_{n} - \sigma_{n-1} \tau_{n+1} 
  & = & 0 
\\ 
  i D_{y} \, \tau_{n} \cdot \rho_{n} - \tau_{n-1} \rho_{n+1} 
  & = & 0 
\\ 
  i D_{y} \, \tau_{n+1} \cdot \tau_{n} - \sigma_{n} \rho_{n+1} 
  & = & 0. 
\label{eq-bilin-6}
\end{eqnarray}
Equations (\ref{eq-bilin-1})--(\ref{eq-bilin-6}) 
belong to the AL hierarchy that has been introduced
in \cite{AL1975}. Indeed, in terms of the functions 
\begin{equation}
  q_{n} = \frac{ \sigma_{n} }{ \tau_{n} }, 
  \qquad
  r_{n} = \frac{ \rho_{n} }{ \tau_{n} }
\end{equation}
equations (\ref{eq-bilin-1})--(\ref{eq-bilin-6}) together with 
(\ref{tau-restiction}) become
\begin{equation}
  \left\{
  \begin{array}{rcl}
	i\partial_{x} q_{n} = \left( 1 - q_{n}r_{n} \right) q_{n+1} 
	\\
	-i\partial_{x} r_{n} = \left( 1 - q_{n}r_{n} \right) r_{n-1} 
  \end{array}
  \right.
\end{equation}
and
\begin{equation}
  \left\{
  \begin{array}{rcl}
	i\partial_{y} q_{n} = \left( 1 - q_{n}r_{n} \right) q_{n-1} 
	\\
	-i\partial_{y} r_{n} = \left( 1 - q_{n}r_{n} \right) r_{n+1} 
  \end{array}
  \right.,
\end{equation}
equations that describe the simplest flows of the AL hierarchy and that can 
be viewed as superposition of the famous discrete NLS and 
discrete modified KdV equations. 

That means that one can use the already known solutions of the AL hierarchy to 
construct the ones for the MRT equation and hence for the FLE.

\section{Dark soliton solutions for the FLE. \label{sec-solitons}}

In the appendix one can find dark-soliton solutions of the generalized AL 
equations \cite{V2002}, the system which is a generalization of 
(\ref{eq-bilin-1})--(\ref{eq-bilin-6}). 
Using these results solutions of the latter can be written as 
\begin{equation}
	\tau_{n} = \tau^{0}_{n} , \qquad
	\sigma_{n} = \tau^{1}_{n} , \qquad
	\rho_{n} = \tau^{-1}_{n} 
\end{equation}
where
\begin{equation}
  \tau^{m}_{n}(x,y)
  =  
  (is)^{m^{2}} r^{n^{2}} \;
  \mu^{m}(x,y)\, \nu^{n}(x,y) \;
  \omega^{m}_{n}(x,y)
\label{ds-tau}
\end{equation}
and
\begin{equation}
  \omega^{m}_{n}(x,y) = 
  \det \left| 
    \mathbb{I} + A(x,y) \mathcal{M}^{m} \mathcal{N}^{n} \right|.
\end{equation}
Here $A(x,y)$ is a square matrix with the elements
\begin{equation}
  A_{jk}(x,y) = 
  \frac{ a_{k}(x,y) }{ L_{j} - R_{k} },
  \qquad
  j,k = 1, ..., N
\label{matrix-a-def}
\end{equation}
\begin{equation}
  \mathcal{M} = \mathop{\mbox{diag}} \left( 
    \frac{ 1 - rL_{j} }{ 1 - rR_{j} } 
  \right),  
  \qquad
  \mathcal{N} = \mathop{\mbox{diag}} \left( 
    \frac{ L_{j} }{ R_{j} } 
  \right),
\end{equation}
$L$ and $R$ are diagonal matrices,
$L = \mathop{\mbox{diag}} \left(L_{j}\right)$ and 
$R = \mathop{\mbox{diag}} \left(R_{j}\right)$, 
related by
\begin{equation}
  \left( L  - \mathbb{I} \right) \left( R  - \mathbb{I} \right) = 
  s^{2} \mathbb{I} 
\end{equation}
($\mathbb{I}$ is the $N \times N$-unit matrix) 
with parameters $r$ and $s$ being related by 
\begin{equation}
  r^{2} - s^{2} = 1.
\end{equation}
The dependence of the solution on the variables $x$ and $y$ is given by 
\begin{equation}
  a_{k}(x,y) = a_{k}^{(0)}
  \exp\left\{ 
  - ir \left( L_{k} - R_{k} \right) x
  - ir \left( L_{k}^{-1} - R_{k}^{-1} \right) y
  \right\},
\end{equation}
\begin{equation}
  \begin{array}{lcl}
  \mu(x,y) & = & \mu^{(0)} \exp\left\{ - i r^{2} \left( x + y \right) \right\}
  \\[2mm]
  \nu(x,y) & = & \nu^{(0)} \exp\left\{ - i s^{2} \left( x - y \right) \right\},
  \end{array}
\end{equation}
where $a_{k}^{(0)}$, $\mu^{(0)}$ and $\nu^{(0)}$ are arbitrary constants.

Bearing in mind physical applications, one can impose on the matrices 
$L$ and $R$ some restrictions which lead to the involution
\begin{equation}
  v_{n} = - \overline{u_{-n}}.
\label{mrte-invol}
\end{equation}
Note that this involution, which is 'natural' for the MRT equations, 
is different from a more usual one for the AL system 
($r_{n} = - \overline{q_{n}}$). However, the former is sufficient to derive 
the 'physical' solutions for the FLE, $v = - \overline{u}$, which is the main 
goal of the present paper. Thus, let us rewrite (\ref{ds-tau}) imposing the 
restrictions
\begin{equation}
  r > 1
\end{equation}
(which implies reality of $s$), 
\begin{equation}
  \begin{array}{lcl}
  L_{k} & = & r + s \exp\left( i\vartheta_{k} \right), 
  \\[2mm]
  R_{k} & = & r + s \exp\left( -i\vartheta_{k} \right), 
  \end{array}
\end{equation}
$\left| \nu^{(0)} \right| = r \left| \mu^{(0)} \right|$, 
and introducing the real matrix $C$,
\begin{equation}
  C = A \, \mathcal{N}^{-1/2}.
\end{equation}
This leads, after some trivial modifications of the above formulae, 
to the following result for the functions $u_{n}$ and $v_{n}$ given 
by (\ref{uv-bilin}):
\begin{eqnarray}
  u_{n}(x,y) & = & 
  is \, e^{-i\varphi(x,y)} r^{-2n}
  \frac{ \det\left| 
    \delta_{jk} + C_{jk}(x,y) e^{ i \left( 2n\beta_{k} + \alpha_{k} \right) } 
  \right|_{j,k=1}^{N}} 
  { \det\left| 
    \delta_{jk} + C_{jk}(x,y) e^{ i \left( 2n+1 \right)\beta_{k} } 
  \right|_{j,k=1}^{N}} 
\label{mrte-ds-u}
  \\[2mm]
  v_{n}(x,y) & = & 
  is \, e^{i\varphi(x,y)} r^{2n}
  \frac{ \det\left| 
    \delta_{jk} + C_{jk}(x,y) e^{ i \left( 2n\beta_{k} - \alpha_{k} \right) } 
  \right|_{j,k=1}^{N}} 
  { \det\left| 
    \delta_{jk} + C_{jk}(x,y) e^{ i \left( 2n-1 \right)\beta_{k} } 
  \right|_{j,k=1}^{N}} 
\label{mrte-ds-v}
\end{eqnarray}
Here $\delta_{jk}$ is the Kronecker's delta, 
\begin{equation}
	\varphi(x,y) = x + \left( 2r^{2} - 1 \right)y + \varphi^{(0)},
\end{equation}
$\varphi^{(0)}$ is an arbitrary real constant,
\begin{equation}
	C_{jk}(x,y) = 
	\frac{ c^{(0)}_{k} \exp\left[ 
    2r \left( \lambda_{k}x - \lambda_{k}^{-1} y \right) \sin\beta_{k} 
    \right]} 
	{ \sin\left( 
	    \displaystyle\frac{ \vartheta_{j}+\vartheta_{k} }{ 2} 
	  \right) }
\end{equation}
where $c^{(0)}_{k}$ are arbitrary real constants,
\begin{equation}
	\lambda_{k} = \left| r + s e^{ i \vartheta_{k} } \right|
\end{equation}
and
\begin{equation}
  \begin{array}{lcl}
  \alpha_{k} & = & 2 \arg\left( s + r e^{ i \vartheta_{k}}  \right) - \beta_{k},
  \\[2mm]
  \beta_{k}  & = & \arg\left( r + s e^{ i \vartheta_{k}}  \right). 
  \end{array}
\end{equation}
Comparing (\ref{mrte-ds-u}) and (\ref{mrte-ds-v}) it is easy to see that they 
indeed satisfy (\ref{mrte-invol}). Finally, setting $n=0$ one comes to 
the dark-soliton solutions of the FLE:
\begin{equation}
  u(x,y) = 
  is \, e^{-i\varphi(x,y)} \, 
  \frac{ \det\left| \delta_{jk} + C_{jk}(x,y) e^{ i \alpha_{k} } \right|_{j,k=1}^{N}} 
       { \det\left| \delta_{jk} + C_{jk}(x,y) e^{ i \beta_{k}  } \right|_{j,k=1}^{N}} 
\end{equation}
and
\begin{equation}
  v(x,y) = - \overline{ u(x,y) }.
\end{equation}

\section{Intersections of hierarchies. \label{sec-intersections}}

To conclude this paper I would like to discuss the question of 'intersections' 
of different integrable hierarchies that one can find in the previous sections: 
the DNLS, AKNS, MRT and AL hierarchies.

The FLE belongs, as has been shown in \cite{F1995}, to the DNLS hierarchy, 
describing its first negative flow. 
Equations (\ref{MRT-eq-1}) and (\ref{MRT-eq-2}) belong to the MRT hierarchy,  
which, in its turn, can be viewed as a discretization of the AKNS hierarchy. 
Finally, the bilinearization of the FLE and MRT equations 
(\ref{MRT-eq-1}) and (\ref{MRT-eq-2}) has led us to the AL hierarchy. 
To summarize, in this paper we were dealing with four integrable models: 
DNLS, NLS, MRT and AL systems. Now I would like to present some explicit 
formulae expressing solutions of the first three equations/hierarchies in 
terms of solutions of the AL model, which is known to be rather 'universal' 
and to 'contain' inside many other integrable models (see lecture \cite{V1998} 
or original papers \cite{V2002,V1995,V1996}). 
To do this I need more equations from the AL hierarchy (its second 
positive flow) which can be presented in the bilinear form as
\begin{equation}
  \begin{array}{lcl}
  D_{t} \, \sigma_{n} \cdot \tau_{n} & = &
  D_{x} \, \sigma_{n+1} \cdot \tau_{n-1} 
  \\[2mm]  
  D_{t} \, \tau_{n} \cdot \rho_{n} & = &
  D_{x} \, \tau_{n+1} \cdot \rho_{n-1} 
  \\[2mm]  
  D_{t} \, \tau_{n+1} \cdot \tau_{n} & = &
  D_{x} \, \rho_{n} \cdot \sigma_{n+1}. 
  \end{array}
\label{alh-second}
\end{equation}
These equations are compatible with 
(\ref{eq-bilin-1})--(\ref{eq-bilin-6}) and (\ref{tau-restiction}) 
being the members of the same hierarchy. Thus the system consisting of 
(\ref{eq-bilin-1})--(\ref{eq-bilin-6}), (\ref{alh-second}) and 
(\ref{tau-restiction}) has nontrivial solutions and one can prove by direct 
calculations the following results:

\begin{itemize}
\item 
For any $n$ the functions 
\begin{equation}
  U = 
  \frac{ \sigma_{n} }{ \tau_{n} },
  \hspace{10mm}
  V = 
  \frac{ \rho_{n-1} }{ \tau_{n-1} },
  \hspace{20mm}
  n = \mbox{constant}
\label{dnls-qr}
\end{equation}
solve the DNLS equation
\begin{equation}
  \left\{
  \begin{array}{rcl}
  i U_{t} + U_{xx} + 2i UV \, U_{x} & = & 0
  \\[2mm]
  -i V_{t} + V_{xx} - 2i UV \, V_{x} & = & 0
  \end{array}
  \right.
\end{equation}
%
\item 
For any $n$ the functions 
\begin{equation}
  Q = 
  \frac{ \sigma_{n+1} }{ \tau_{n} },
  \hspace{10mm}
  R = 
  \frac{ \rho_{n-1} }{ \tau_{n} },
  \hspace{20mm}
  n = \mbox{constant}
\label{nls-qr}
\end{equation}
solve the NLS equation
\begin{equation}
  \left\{
  \begin{array}{rcl}
    iQ_{t} + Q_{xx} + 2 Q^{2} R & = & 0 
  \\[2mm]
  - iR_{t} + R_{xx} + 2 Q R^{2} & = & 0 
  \end{array}
  \right.
\end{equation}
%
\item 
The functions 
\begin{equation}
  u_{n} = 
  \frac{ \sigma_{n-1} }{ \tau_{n} },
  \hspace{10mm}
  v_{n} = 
  \frac{ \rho_{n} }{ \tau_{n-1} }
\label{mrt-qr}
\end{equation}
solve the MRT equations (\ref{MRT-eq-1}) and (\ref{MRT-eq-2}).
\end{itemize}

\noindent
These results can be extended to the level of hierarchies:

\begin{itemize}
\item 
For any $n$ the functions $U$ and $V$ given by (\ref{dnls-qr})
where $\rho_{n}$, $\sigma_{n}$ and $\tau_{n}$ are solutions of the AL hierarchy 
solve all equations of the DNLS hierarchy.
\item 
For any $n$ the functions $Q$ and $R$ given by (\ref{nls-qr})
where $\rho_{n}$, $\sigma_{n}$ and $\tau_{n}$ are solutions of the AL hierarchy 
solve all equations of the AKNS hierarchy.
\item 
The functions $u_{n}$ and $v_{n}$ given by (\ref{mrt-qr})
where $\rho_{n}$, $\sigma_{n}$ and $\tau_{n}$ are solutions of the AL hierarchy 
solve all equations of the MRT hierarchy.

\end{itemize}

\section*{Acknowledgements.}

This work has been partially supported by grants FIS2007-29093-E  
(Ministerio de Educaci\'on y Ciencia, Spain) 
and PCI-08-0093 (Consejer\'ia de Educaci\'on y Ciencia, Junta de Comunidades
de Castilla-La Mancha, Spain).

\appendix
\section{Dark solitons of the AL equations}

The dark solitons for the AL equations were obtained in \cite{VK1992} using 
the inverse scattering method. Here I give another, purely algebraic, 
derivation of these solutions which is based on the Fay-like identities 
for the determinants of some special matrices.

To make the following formulae more readable and to expose the inner structure 
of equations (\ref{eq-bilin-1})--(\ref{eq-bilin-6}) and (\ref{tau-restiction}) 
I rewrite them in terms of the double infinite set of tau-functions, 
$\tau^{m}_{n}$, given by
\begin{equation}
	\tau^{0}_{n} = \tau_{n}, \qquad
	\tau^{1}_{n} = \sigma_{n}, \qquad
	\tau^{-1}_{n} = \rho_{n}
\end{equation}
and
\begin{equation}
  \begin{array}{rcl}
	\tau^{m+1}_{n} & = & 
	  \displaystyle\frac{ 1 }{ \tau^{m-1}_{n} }
	  \left[ 
	    \left(\tau^{m}_{n}\right)^{2} - \tau^{m}_{n-1}\tau^{m}_{n+1}
	  \right]
	\\[2mm]
	\tau^{-(m+1)}_{n} & = & 
	  \displaystyle\frac{ 1 }{ \tau^{-(m-1)}_{n} }
	  \left[ 
	    \left(\tau^{-m}_{n}\right)^{2} - \tau^{-m}_{n-1}\tau^{-m}_{n+1}
	  \right]
\end{array}
\qquad
  m = 1, 2, ...
\end{equation}
In terms of the new tau-functions equations
(\ref{eq-bilin-1})--(\ref{eq-bilin-6}) become
\begin{eqnarray}
  i D_{x} \, \tau^{m+1}_{n} \cdot \tau^{m}_{n} & = & 
  \phantom{-} \tau^{m}_{n-1} \tau^{m+1}_{n+1} 
\label{ale-pos-m}
\\ 
  i D_{x} \, \tau^{m}_{n+1} \cdot \tau^{m}_{n} & = & 
  - \tau^{m-1}_{n} \tau^{m+1}_{n+1} 
\label{ale-pos-n}
\\[3mm] 
  i D_{y} \, \tau^{m+1}_{n} \cdot \tau^{m}_{n} & = & 
  \phantom{-} \tau^{m}_{n+1} \tau^{m+1}_{n-1} 
\label{ale-neg-m}
\\ 
  i D_{y} \, \tau^{m}_{n+1} \cdot \tau^{m}_{n} & = & 
  \phantom{-} \tau^{m-1}_{n+1} \tau^{m+1}_{n} 
\label{ale-neg-n}
\end{eqnarray}
subjected to the restriction (\ref{tau-restiction})
\begin{equation}
  \left(\tau^{m}_{n}\right)^{2} = 
  \tau^{m-1}_{n} \tau^{m+1}_{n}
	+ 
	\tau^{m}_{n-1}\tau^{m}_{n+1}.
\label{ale-restriction}
\end{equation}

The dark-soliton solutions of the AL hierarchy are made of the 
$N \times N$ matrices $A$ that satisfy the equation 
\begin{equation}
  L A - A R = | \,\ell\, \rangle \langle a |
\end{equation}
where $L$ and $R$ are constant diagonal matrices, $| \,\ell\, \rangle$ is 
constant $N$-column, 
$| \,\ell\, \rangle = \left( \ell_{1}, ... , \ell_{N} \right)^{T}$
and $\langle a |$ is $N$-row depending on the coordinates 
describing the AL flows,
$\langle a(x,y) | = \left( a_{1}(x,y), ... , a_{N}(x,y) \right)$. 
By straightforward algebra it can be shown that 
the determinants 
\begin{equation}
  \omega\left( A \right) = \det \left| \mathbb{I} + A \right|
\label{omega-def}
\end{equation}
satisfy the Fay-like identity
\begin{equation}
  \alpha ( \beta - \gamma ) \; 
  \omega_{\alpha}\omega_{\beta\gamma} 
  +
  \beta ( \gamma - \alpha ) \; 
  \omega_{\beta}\omega_{\alpha\gamma} 
  +
  \gamma ( \alpha - \beta ) \; 
  \omega_{\gamma}\omega_{\alpha\beta} 
  =  0
\label{fay}
\end{equation}
where
\begin{equation}
  \omega = \omega\left( A \right),
  \qquad
  \omega_{\zeta} = \omega\left( A K_{\zeta} \right),
  \qquad
  \omega_{\xi\eta} = \omega\left( A K_{\xi}K_{\eta} \right)
\end{equation}
with
\begin{equation}
  K_{\zeta} = 
  \left( \mathbb{I} - \zeta L \right)
  \left( \mathbb{I} - \zeta R \right)^{-1}.
\end{equation}
In what follwos I use the doubly infinite family of matrices 
$A^{m}_{n}$ defined by 
\begin{equation}
  A^{m+1}_{n} = A^{m}_{n} \mathcal{M}, 
  \qquad
  A^{m}_{n+1} = A^{m}_{n} \mathcal{N} 
\end{equation}
where
\begin{equation}
  \mathcal{M} = K_{r}, 
  \qquad
  \mathcal{N} = K_{\infty} 
\end{equation}
and relate the matrices $L$ and $R$ by
\begin{equation}
  \left( L - r\mathbb{I} \right)
  \left( R - r\mathbb{I} \right) = 
  \left( r^{2} - 1 \right) \mathbb{I} 
\end{equation}
which leads to
\begin{equation}
  K_{r} K_{\infty} = K_{1/r}. 
\end{equation}
Now equation (\ref{fay}) with 
$(\alpha, \beta, \gamma) = (r, 1/r, \infty)$ 
can be rewritten as
\begin{equation}
  \left(\omega^{m}_{n} \right)^{2} = 
	\rho^{2} \; \omega^{m-1}_{n} \; \omega^{m+1}_{n}
	+ r^{2} \; \omega^{m}_{n-1} \; \omega^{m}_{n+1}
\label{omega-restr}
\end{equation}
where 
\begin{equation}
  \omega^{m}_{n} = \omega\left( A^{m}_{n} \right)
\end{equation}
and $\rho$ is given by $\rho = \sqrt{1 - r^{2}}$.
In a similar way, (\ref{fay}) with $(\beta,\gamma)$ equal to
$(1/r,\infty)$, $(r,1/r)$ and $(r,\infty)$ leads to the following 
identities: 
\begin{eqnarray}
&&
	( r\alpha - 1 ) \; \omega^{m}_{n} \; \tilde\omega^{m+1}_{n}
	+ \omega^{m+1}_{n} \; \tilde\omega^{m}_{n}
	= r\alpha \; \omega^{m+1}_{n+1} \; \tilde\omega^{m}_{n-1}
\label{alh-fay-1}
\\&&
	( \alpha - r ) \; \omega^{m}_{n+1} \; \tilde\omega^{m}_{n}
	+ r ( 1 - r\alpha  ) \; \omega^{m}_{n} \; \tilde\omega^{m}_{n+1}
	= \alpha \rho^{2} \; \omega^{m+1}_{n+1} \; \tilde\omega^{m-1}_{n}
\label{alh-fay-2}
\\&&
	( \alpha - r ) \; \omega^{m}_{n+1} \; \tilde\omega^{m+1}_{n}
	+ r \; \omega^{m+1}_{n} \; \tilde\omega^{m}_{n+1}
	= \alpha \; \omega^{m+1}_{n+1} \; \tilde\omega^{m}_{n}.
\label{alh-fay-3}
\end{eqnarray}
with
\begin{equation}
  \tilde\omega^{m}_{n} = \omega\left( A^{m}_{n}K_{\alpha} \right).
\end{equation}

Analysing (\ref{alh-fay-1})--(\ref{alh-fay-3}) with 
$\alpha = \lambda(\xi)$ and $\alpha = \bar\lambda(\eta)$ where
\begin{equation}
  \xi = \frac{ \lambda(\xi) \left[ r\lambda(\xi) - 1 \right] }
             { \lambda(\xi) - r } 
\end{equation}
and 
\begin{equation}
  \eta = 
  \frac{ \bar\lambda(\eta) - r  }
       { \bar\lambda(\eta) \left[ r\bar\lambda(\eta) - 1 \right] } 
\end{equation}
(these equations play the role of the dispersion laws of the positive and 
negative AL subhierarchies under finite-density boundary conditions) 
one can conclude, after identifying the Miwa's shifts with the action of the 
$K_{\lambda(\xi)} K_{1/r}^{-1}$ and $K_{\bar\lambda(\eta)} K_{r}^{-1}$, 
that it is possible to introduce the dependence of $A^{m}_{n}$ on an infinite 
number of 'times' in such a way that $\omega^{m}_{n}$ become solutions of the 
functional equations closely related to the so-called functional representation 
of the AL hierarchy \cite{V2002}. However, I do not discuss here these equations in 
details because my purpose is to solve only the simplest AL equations 
(\ref{ale-pos-m})--(\ref{ale-neg-n}) which can be done as follows.

Let the $x$-dependence of $A^{m}_{n}$ being determined by 
\begin{equation}
  i \partial_{x} A^{m}_{n} = A^{m}_{n} \, X
\end{equation}
where
\begin{equation}
  X = 
  \lim_{\xi \to 0} \frac{ 1 }{ \xi }
  \left[ K_{\lambda(\xi)} K_{1/r}^{-1} - \mathbb{I} \right]
  =
  r \left( L - R \right).
\end{equation}
Replacing in $\alpha$ (\ref{alh-fay-1})--(\ref{alh-fay-2}) with $\lambda(\xi)$ 
and taking the $\xi \to 0$ limit one comes to 
\begin{eqnarray}
	i D_{x} \; \omega^{m+1}_{n} \cdot \omega^{m}_{n} & = & 
	r^{2} \left[ 
	  \omega^{m}_{n-1} \; \omega^{m+1}_{n+1} - \omega^{m}_{n} \; \omega^{m+1}_{n}
  \right]
\label{omega-pos-m}
\\[2mm]
	i D_{x} \; \omega^{m}_{n+1} \cdot \omega^{m}_{n} & = & 
	\rho^{2} \left[ 
	  \omega^{m}_{n} \; \omega^{m}_{n+1} - \omega^{m-1}_{n} \; \omega^{m+1}_{n+1}
  \right].
\label{omega-pos-n}
\end{eqnarray}
In a similar way, introducing the $y$-dependence of $A^{m}_{n}$ by 
\begin{equation}
  i \partial_{y} A^{m}_{n} = A^{m}_{n} \, Y 
\end{equation}
with
\begin{equation}
  Y = 
  \lim_{\eta \to 0} \frac{ 1 }{ \eta }
  \left[ K_{\bar\lambda(\eta)} K_{r}^{-1} - \mathbb{I} \right]
  = 
  r \left( L^{-1} - R^{-1} \right)
\end{equation}
one can get, after taking the $\eta \to 0$ limit in 
(\ref{alh-fay-2})--(\ref{alh-fay-3}) with $\alpha = \bar\lambda(\eta)$, 
\begin{eqnarray}
	i D_{y} \; \omega^{m+1}_{n} \cdot \omega^{m}_{n} & = & 
	r^{2} \left[ 
	  \omega^{m+1}_{n-1} \; \omega^{m}_{n+1} - \omega^{m}_{n} \; \omega^{m+1}_{n}
  \right]
\label{omega-neg-m}
\\[2mm]
	i D_{y} \; \omega^{m}_{n+1} \cdot \omega^{m}_{n} & = & 
	\rho^{2} \left[ 
	  \omega^{m+1}_{n} \; \omega^{m-1}_{n+1} - \omega^{m}_{n} \; \omega^{m}_{n+1}
  \right].
\label{omega-neg-n}
\end{eqnarray}
Now one can obtain 
solutions of the AL equations by simple
transformation
\begin{equation}
  \tau^{m}_{n}
  =  
  \rho^{m^{2}} r^{n^{2}} \;
  \mu^{m}\, \nu^{n} \;
  \omega^{m}_{n}.
\end{equation}
It is easy to show starting from (\ref{omega-restr}) that
$\rho^{m^{2}} r^{n^{2}}$ factor ensures that tau-functions $\tau^{m}_{n}$ 
satisfy (\ref{ale-restriction}). 
Then, by simple calculations one can get that if 
$\mu$ and $\nu$ depend on $x$ and $y$ as
\begin{equation}
  \begin{array}{lcr}
  \mu & = & \mu^{(0)} \exp\left\{ -ir^{2} (y+x) \right\}
  \\
  \nu & = & \nu^{(0)} \exp\left\{ -i\rho^{2} (y-x) \right\}
  \end{array}
\end{equation}
then equations (\ref{omega-pos-m})--(\ref{omega-neg-n}) imply
that $\tau^{m}_{n}$ solve AL equations 
(\ref{ale-pos-m})--(\ref{ale-neg-n}).

Noting that determinants (\ref{omega-def}) are invariant under 
transformations $A \to U^{-1}AU$ one can eliminate (without loss of 
generality the constants $\ell_{j}$ ($\ell_{j} \to 1$) 
by redefining the functions $a_{k}(x,y)$
($a_{k}(x,y) \to a_{k}(x,y) \ell_{k}$)
coming to the expressions for the elements of the matrix $A$ 
(\ref{matrix-a-def}). Also, considering solutions of the MRT equations 
and the FLE it is convenient to use instead of $\rho$ another parameter,
$s$, given by $\rho = is$ (see section \ref{sec-solitons}).

\section*{References}                                                         %

\end{document}